\documentclass[fleqn,usenatbib]{mnras}

\usepackage{newtxtext,newtxmath}

\usepackage[T1]{fontenc}

\DeclareRobustCommand{\VAN}[3]{#2}
\let\VANthebibliography\thebibliography
\def\thebibliography{\DeclareRobustCommand{\VAN}[3]{##3}\VANthebibliography}

\usepackage{graphicx}	
\usepackage{amsmath}	
\usepackage{amssymb}	
\usepackage[T1]{fontenc}
\usepackage{ae,aecompl}
\usepackage{epstopdf}

\newcommand{\Msun}{\mbox{$M_{\odot}$}}

\newcommand{\Mwd}{\mbox{$M_{\mathrm{WD}}$}}
\newcommand{\Rwd}{\mbox{$R_{\mathrm{WD}}$}}
\newcommand{\Teff}{\mbox{$T_{\mathrm{eff}}$}}

\newcommand{\kms}{km/s}

\newcommand{\gppr}{\stackrel{>}{\scriptstyle \sim}}
\newcommand{\gappr}{\raisebox{-0.4ex}{$\gppr$}}
\newcommand{\lppr}{\stackrel{<}{\scriptstyle \sim}}
\newcommand{\lappr}{\raisebox{-0.4ex}{$\lppr$}}
\newcommand{\tyc}{TYC\,7218}
\newcommand{\tycc}{TYC\,7218 }
\usepackage[usenames,dvipsnames]{xcolor}
\usepackage{color,colortbl}
\usepackage{soul}
\usepackage{subfig}

\title[The White Dwarf Binary Pathways Survey III]{The White Dwarf Binary Pathways Survey III: 
contamination from hierarchical triples containing a white dwarf}
\author[F. Lagos et al.]
{
F. Lagos,$^{1,2,13}$\thanks{felipe.lagos@postgrado.uv.cl}
M.R. Schreiber,$^{1,2}$
S.G. Parsons,$^{3}$
A. Zurlo,$^{4}$
D. Mesa,$^{14}$
B.T. G\"ansicke,$^{5,6}$
\newauthor
R. Brahm,$^{7,8,15}$ 
C. Caceres,$^{2,9}$
H. Canovas,$^{10}$
M-S. Hernandez,$^1$
A. Jordan,$^{11,8}$ 
\newauthor
D. Koester,$^{12}$
L. Schmidtobreick,$^{13}$
C. Tappert,$^{1}$
M. Zorotovic,$^{1}$
\\
$^1$Instituto de F{\'i}sica y Astronom{\'i}a, Universidad de Valpara{\'i}so, Valpara{\'i}so, Chile\\
$^2$Millennium Nucleus for Planet Formation, NPF, Universidad de Valpara{\'i}so, Valpara{\'i}so, Chile\\
$^3$Department of Physics and Astronomy, University of Sheffield, Sheffield S3 7RH, UK\\
$^4$Nucleo de Astronomia, Facultad de Ingenieria y Ciencias, Universidad Diego Portales, Av. Ejercito 441, Santiago, Chile\\
$^5$Department of Physics, University of Warwick, Coventry CV4 7AL, UK\\
$^6$Centre for Exoplanets and Habitability, University of Warwick, Coventry CV4 7AL, UK\\
$^7$Instituto de Astrofsica, Facultad de Fsica, Pontificia Universidad Catolica de Chile, Av. Vicuna Mackenna 4860, 7820436 Macul, Santiago, Chile\\
$^8$Millennium Institute of Astrophysics, Chile\\
$^9$Departamento de Ciencias Fisicas, Facultad de Ciencias Exactas, Universidad Andres Bello, Av. Fernandez Concha 700, Las Condes, Santiago, Chile\\
$^{10}$Aurora Technology B.V. for ESA, ESA-ESAC, Camino Bajo del Castillo s/n,28692 Villanueva de la Ca\~{n}ada, Madrid,Spain \\
$^{11}$Facultad de Ingeniera y Ciencias, Universidad Adolfo Ibanez, Santiago, Chile \\
$^{12}$Institut f\"ur Theoretische Physik und Astrophysik, Universit\"at Kiel, 24098 Kiel, Germany\\
$^{13}$European Southern Observatory (ESO), Alonso de Cordova 3107, Vitacura, Santiago, Chile\\
$^{14}$INAF - Osservatorio Astronomico di Padova, Italy\\
$^{15}$Center of Astro-Engineering UC, Pontificia Universidad Cat\'olica de Chile, Av. Vicu\~{n}a Mackenna 4860, 7820436 Macul, Santiago, Chile
}

\date{Accepted XXX. Received YYY; in original form ZZZ}

\pubyear{2020}

\begin{document}
\label{firstpage}
\pagerange{\pageref{firstpage}--\pageref{lastpage}}

\maketitle
\begin{abstract}
\noindent
The \textit{White Dwarf Binary Pathways Survey} aims at increasing the number of known detached A, F, G and K main sequence stars in close orbits with white dwarf companions (WD+AFGK binaries) to refine our understanding about compact binary evolution and the nature of Supernova Ia progenitors.
These close WD+AFGK binary stars are expected to form through common envelope evolution, in which tidal forces tend to circularize the orbit.
However, some of the identified WD+AFGK binary candidates show eccentric orbits, indicating that these systems are either formed through a different mechanism or perhaps they are not close WD+AFGK binaries. We observed one of these eccentric WD+AFGK binaries with SPHERE and find that the system TYC\,7218-934-1 is in fact a triple system where the WD is a distant companion. 
The inner binary likely consists of the G-type star plus an unseen low mass companion in an eccentric orbit. 
Based on this finding, we estimate the fraction of triple systems that could contaminate the WD+AFGK sample. We find that less than 15 per cent of our targets with orbital periods shorter than 100 days might be hierarchical triples. 
\end{abstract}




\section{Introduction}

The unique potential of Type Ia 
Supernovae (SN Ia) 
as distance indicators, 
sufficiently bright to serve as yardsticks on 
cosmological 
distance scales \citep{branch+tammann92-1}, 
has made them some of the most 
important objects for our understanding in the Universe and has led to 
the discovery of its accelerating 
expansion
\citep{riessetal98-1,perlmutteretal99-1}. Despite its importance, it is not clear which evolutionary pathways lead to SN\,Ia explosions, and the SN\,Ia progenitor puzzle remains a crucial unsolved astronomical question. 

The two main formation channels for SN Ia explosions, i.e., the single \citep{webbink84-1} and double \citep{iben+tutukov84-1} degenerate channels, involve thermonuclear explosions of WDs close to the Chandrasekhar mass limit. However, in  recent years it has become evident that reaching a mass close to the Chandrasekhar limit might not be a strict criterion for SN\,Ia explosions, enlarging the possible evolutionary scenarios towards single or double degenerate SN\,Ia detonations  \citep{finketal07-1,simetal10-1,kromeretal10-1,brooksetal16-1,guillochonetal10-1,vankerkwijketal10-1,pakmoretal13-1}. The situation has become even more complex as recently suggested evolutionary channels such as the core-degenerate \citep{livio+riess03-1,sokeretal13-1,aznar-siguanetal15-1} and the WD collisional scenario \citep{raskinetal09-1,rosswogetal09-1} represent possible alternatives to the classical single and double degenerate channels.

In spite of the growing variety of potential SN\,Ia progenitor systems, it is clear 
that the progenitors are close binaries that contain
at least one WD. However, it remains an open question how these 
close binaries form and how frequently 
a given SN\,Ia progenitor configuration is 
produced.  
Usually the initial main 
sequence binary, consisting of 
two stars with masses exceeding $1\Msun$, 
is assumed to evolve through a common envelope phase
\citep[e.g.][]{webbink84-1,zorotovicetal10-1} 
to generate the small binary separations required 
for any interaction between the stars. The close binary emerging from the common envelope phase, i.e. 
the post common envelope binary (PCEB), consists of 
a WD and a main sequence companion star. 
The orbital separation of a recently formed PCEB largely determines the future evolution of the system. Broadly speaking, if the orbital period 
is longer than 1-2 days, a second phase of mass transfer will be initiated when the secondary star evolves off the main sequence which, depending on the mass ratio of the system, may lead to a second common envelope phase and potentially to a double degenerate binary. If the orbital periods are shorter than 1-2 days, angular momentum loss can drive the system into mass transfer when the secondary star is still on the main sequence, which, again depending on the mass ratio of the binary system, can cause thermally unstable mass transfer and stable nuclear burning on the surface of the WD. Such systems, called Super Soft X-ray Sources, are considered possible single degenerate SN\,Ia progenitors \citep{parsons15}.

Current binary population models rely on simple equations for common envelope evolution and, unfortunately, the common envelope efficiency is only well constrained for low-mass companions \citep{zorotovicetal10-1,2011A&A...536A..43N} which are irrelevant for SN\,Ia progenitor studies.   
It is therefore currently impossible to use binary population models to reliably predict relative numbers of SN\,Ia explosions generated by the single or the double degenerate channels. 

To progress with this situation, we began a large scale survey of PCEBs, consisting of main sequence stars of spectral type AFGK plus a WD. As the WD is outshined by the AFGK secondary stars at optical wavelengths, we combine spectroscopic surveys with the Galaxy Evolution Explorer (\textit{GALEX}) database to identify AFGK stars with UV excess indicative of a WD companion \citep{parsons16,rebassa-mansergas17}. We then 
use radial velocity measurements to identify the close binary systems and to measure their periods. So far, we have identified the first pre-supersoft X-ray binary
system \citep{parsons15}, confirmed that our target selection is reliable and contains few contaminants \citep{parsons16}, and have published the first results of our radial velocity campaign \citep{rebassa-mansergas17}. 

One of our close WD+AFGK candidate stars identified by matching data from \textit{GALEX} and the Radial Velocity Experiment (RAVE) survey (\citealt{parsons16}) was the G-star TYC\,7218-934-1 (henceforth \tyc). We here present radial velocity measurements of this object and find a short orbital period of 13.6\,days, with an unexpected
high eccentricity of \textit{e}=0.46. This eccentricity excludes CE evolution for the origin of the current short period. 
Investigating the origin of the eccentric orbit using high-contrast imaging, we find that \tycc is in fact a 
hierarchical triple star system with the WD being the tertiary component. 
We estimate that such triple star systems contaminate our 
PCEB sample by $\lappr15$ per cent.

\section{\tycc as a target of the SN\,Ia pathway project}

As described in detail in \citet{parsons16}, we identify WD+AFGK candidate binaries using UV excesses found by cross-matching optical surveys with \textit{GALEX} data. \tycc was one of our \textit{RAVE} targets where we found clear evidence for a UV excess in \textit{GALEX}. We consequently followed up this object with high resolution spectroscopy to determine the orbital period of the binary and to obtain more information on the properties of its stellar components.

\subsection{Characterising the G-type star}

We obtained high resolution spectroscopy of \tycc with the echelle spectrograph (resolving power R$\simeq$40,000; wavelength range $\sim$3700{\AA} to $\sim$7000\AA) on the 2.5-m Du Pont telescope located at Las Campanas Observatory, Chile and with FEROS (R$\sim$48,000; $\sim$3500{\AA} to $\sim$9200\AA) on the 2.2-m Telescope at La Silla, Chile. 
Data obtained with both of these
spectrographs were extracted and analysed using the Collection of Elemental
Routines for Echelle Spectra ({\sc ceres}) package \citep{brahm17}, which was
developed to process spectra coming from different instruments in an
homogeneous and robust manner. After performing standard image reductions,
spectra were optimally extracted following \citet{marsh89-1} and calibrated in
wavelength using reference ThAr Lamps. For FEROS data, the instrumental drift
in wavelength through the night was corrected with a secondary fiber observing
a ThAr lamp. In the case of the Du Pont data, ThAr spectra were acquired
before and after each science observation. Wavelength solutions were shifted
to the barycenter of the solar system. 

We estimated the stellar parameters for the main-sequence star in
\tycc by comparing the observed data against a synthetic grid of
stellar spectra \citep{coelho05-1}. The synthetic spectra were degraded to the
resolution of the Du Pont echelle and FEROS by convolving them with a Gaussian. The optimal fit for each spectrum was found by chi-square
minimisation. We then combined the results from each spectral fit and used the
average values, yielding an effective temperature $T_\mathrm{eff}=5790\pm50$\,K, surface gravity $\log{g}=4.51\pm0.05$
(in cgs units) and metallicity [Fe/H]=$0.00\pm0.05$. Using the Torres relation
\citep{torresetal10-1} we find a mass and radius of $1.04\pm0.02$M$_\odot$ and
$1.06\pm0.06$R$_\odot$ for the G2V star 
in \tyc.

We also took a medium resolution spectrum (R$\sim$5,000) of \tycc 
with X-shooter \citep{dodoricoetal06-1} mounted at the Cassegrain focus of VLT-UT2
at Paranal on the 11th of May 2015. X-shooter is comprised of three
detectors which permits to obtain simultaneous data from the UV cutoff at
0.3$\mu$m to the $K$-band at 2.4$\mu$m. Our data consisted of three 60 second
exposures, which were reduced using the standard pipeline release of the
X-shooter Common Pipeline Library (CPL) recipes (version 2.5.2). The
instrumental response was removed and the spectrum flux calibrated by
observing the spectrophotometric standard star LTT\,3218. The spectrum 
was not corrected for telluric features. 
The X-shooter spectrum extends our spectroscopic coverage to shorter and longer wavelengths and agrees well with the solution obtained from fitting our FEROS and Du Pont spectra. 

\subsection{The white dwarf component as seen with \textit{HST}\label{sec:coswd}}

The G-type star characterised in the previous section outshines the WD at optical wavelengths. As described in detail in \citet{parsons16}, we spectroscopically observed nine UV-excess objects (WD+AFGK candidates) with \textit{HST} in order to confirm that the excess is due to a WD companion. Depending upon the brightness of 
the target we either used the Space Telescope Imaging Spectrograph (STIS) or the Cosmic Origins Spectrograph (COS). \tycc was among our \textit{HST} targets. 
We obtained two COS spectra of \tycc on 2015 April 29 and May 7, each with an exposure time of 2175\,s. We used the G130M grating centred on $1291\,$\AA, resulting in a wavelength coverage $1130-1430$\,\AA\ at a spectral resolution of $\simeq0.1$\,\AA. We dithered the spectrum in the dispersion direction across all four FP-POS settings to mitigate fixed pattern noise of the COS detector. The data were reduced with CALCOS V3.1.1. 

The broad Ly-$\alpha$ line and the blue continuum confirm that the \textit{UV} excess detected with \textit{GALEX} is due to the presence of a hot compact object: a WD.
In contrast to the close binary TYC 6760-497-1 \citep{parsons15}, no sharp metal lines were detected, hence it was not possible to measure the radial velocity of the WD. 

\citet{parsons16} performed a preliminary analysis of the effective temperature of the WD in \tycc and obtained $\Teff\simeq16\,500$\,K, assuming a fixed WD mass of $\Mwd=0.6\,\Msun$. To improve upon this estimate, we re-analysed the COS spectrum 
making use of the accurate distance to \tycc provided by the \textit{Gaia} Data Release~2 (DR2) parallax measurement \citep{gaiaetal18-1} and a grid of hydrogen model atmospheres spanning $\Teff=14\,000-40\,000$\,K and surface gravities of $\log g=7.5-9.4$ which were computed with the code of \cite{koester10-1}. 

\begin{figure}
\begin{center}
 \includegraphics[width=\columnwidth]{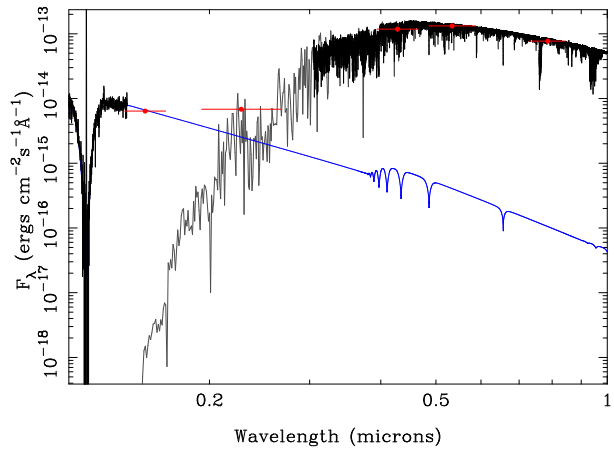}
 \caption{Spectral energy distribution of \tyc. The thick black lines show the
   X-shooter and the HST spectrum, while red dots correspond to photometric measurements. 
   The best fit to the system is obtained with a G2V main
sequence star (thin black line) and a WD with an effective temperature of $\Teff=17\,126\pm150$\,K (blue line).} 
 \label{fig:sedplot}
 \end{center}
\end{figure}

For a fixed $\log g$, a fit to the COS spectrum 
results in \Teff, and using the cooling models of \citealt{holberg+bergeron06-1}, \citealt{kowalski+saumon06-1} and \citealt{tremblayetal11-2}\footnote{http://www.astro.umontreal.ca/$\sim$bergeron/CoolingModels/}, we can calculate the mass and radius of the WD. 
The scaling factor between the model spectrum and the COS data relates 
to the WD radius ($\Rwd$) and distance as 
\begin{equation}
    \frac{F_\mathrm{COS}}{H_\mathrm{mod}}=4\pi\frac{\Rwd^2}{d^2}
\end{equation}
with $F_\mathrm{COS}$ the observed flux and $H_\mathrm{mod}$ the Eddington flux of the model. We then iterate the fit over $\log g$ until the distance implied by the fit agrees with that implied by the \textit{Gaia} DR2 parallax, $d=183.8\pm1.6$\,pc. 
Fitting the two COS spectra individually, and assuming zero reddening, the best-fit parameters obtained are $\Teff=17\,255\pm36$\,K and $\log g=8.222\pm0.015$ (2015 April 29) and $\Teff=17\,106\pm38$\,K and $\log g=8.179\pm0.016$ (2015 May 7). The given uncertainties are purely statistical in nature, and the discrepancy between the two fits indicates that additional systematic uncertainties dominate the results. We adopt $\Teff=17\,126\pm150$\,K and $\log g=8.19\pm0.03$ as the final parameters, obtained from fitting the combined COS spectra and estimating the total uncertainties from the fits to the individual COS observations. 
The WD mass implied by these atmospheric parameters is $M_\mathrm{WD}=0.73\pm0.03\,\mathrm{M}_\odot$. Repeating the previous analysis, but considering a reddening E(B-V)=0.017\footnote{https://stilism.obspm.fr} in the spectral fit, does not lead to significant differences ($\approx 2\%$) in the final parameters of the WD compared to the zero-reddening case. 
The \textit{HST} and X-shooter spectra are shown in Fig. \ref{fig:sedplot} together with the best fit for the WD and the G-type star.

\subsection{The radial velocity curve of \tyc}
\label{radial curve}
The observing strategy of our SN\,Ia pathway project is to
take a few spectra per object in our UV excess target sample to test for
radial velocity variations. If radial velocity variations are detected, we then aim
at measuring the orbital period of the close binary. 
The first two spectra taken of \tycc with the
Echelle spectrograph at the Du Pont  telescope clearly showed small but
significant radial velocity variations (see Table\,\ref{tab:vels}). 
We therefore followed up the system with the Du Pont/Echelle and MPG2.2/FEROS spectrographs until the orbital period was clearly measured. 
\begin{table}
 \centering
  \caption{Radial velocity measurements for the G star in TYC\,7218-934-1.}
  \label{tab:vels}
  \begin{tabular}{@{}lccc@{}}
  \hline
  BJD (TDB)             & Velocity & Uncertainty & Telescope/ \\
  (mid-exposure)   & (\kms)   & (\kms)      & instrument \\
  \hline
  2456809.55612643 &  9.966 & 0.500 & Du Pont/Echelle \\
  2456810.53044180 &  6.760 & 0.500 & Du Pont/Echelle \\
  2456810.64079226 &  5.518 & 0.500 & Du Pont/Echelle \\
  2456811.46154176 &  1.996 & 0.500 & Du Pont/Echelle \\
  2456811.52899051 &  1.177 & 0.500 & Du Pont/Echelle \\
  2456811.63107790 &  0.078 & 0.500 & Du Pont/Echelle \\
  2456827.46756122 & -3.923 & 0.010 & MPG2.2/FEROS \\
  2456827.53633937 & -2.975 & 0.010 & MPG2.2/FEROS \\
  2456828.46953698 & 16.493 & 0.010 & MPG2.2/FEROS \\
  2456828.55232328 & 17.966 & 0.010 & MPG2.2/FEROS \\
  2456829.52614915 & 26.724 & 0.010 & MPG2.2/FEROS \\
  2456829.57929164 & 26.879 & 0.010 & MPG2.2/FEROS \\
  2456829.62074015 & 26.996 & 0.010 & MPG2.2/FEROS \\
  2456830.62692793 & 27.289 & 0.012 & MPG2.2/FEROS \\
  2456831.48604658 & 25.761 & 0.010 & MPG2.2/FEROS \\
  2456832.47118249 & 23.405 & 0.010 & MPG2.2/FEROS \\
  2456832.62146085 & 23.021 & 0.010 & MPG2.2/FEROS \\
  2456833.47375716 & 20.727 & 0.010 & MPG2.2/FEROS \\
  2456834.48096122 & 17.777 & 0.011 & MPG2.2/FEROS \\
  2456834.62990540 & 17.311 & 0.013 & MPG2.2/FEROS \\
  2456835.49422311 & 14.558 & 0.010 & MPG2.2/FEROS \\
  2456835.62537693 & 14.135 & 0.012 & MPG2.2/FEROS \\
  2457000.81838345 &  6.481 & 0.010 & MPG2.2/FEROS \\
  2457001.80957701 &  1.717 & 0.020 & MPG2.2/FEROS \\
  2457001.84002488 &  1.576 & 0.018 & MPG2.2/FEROS \\
  2457002.81549501 & -3.854 & 0.010 & MPG2.2/FEROS \\
  2457003.81567332 & -7.157 & 0.010 & MPG2.2/FEROS \\
  2457003.84959665 & -7.105 & 0.010 & MPG2.2/FEROS \\
  2457004.79505517 &  5.728 & 0.013 & MPG2.2/FEROS \\
  2457004.84905627 &  7.003 & 0.010 & MPG2.2/FEROS \\
  2457386.82926372 & 25.034 & 0.010 & MPG2.2/FEROS \\
  \hline
  \end{tabular}
\end{table}
The radial velocities were computed from the optical echelle spectra using the
cross-correlation technique against a binary mask representative of a G2-type
star. The uncertainties in radial velocity were 
computed using scaling
relations \citep[for more detail see][]{jordan14} with 
the signal-to-noise
ratio and width of the cross-correlation peak, which 
were calibrated with
Monte Carlo simulations. 
The observing dates and the obtained radial velocities are shown in Table\,\ref{tab:vels}. 

\begin{figure}
\begin{center}
 \includegraphics[width=\columnwidth]{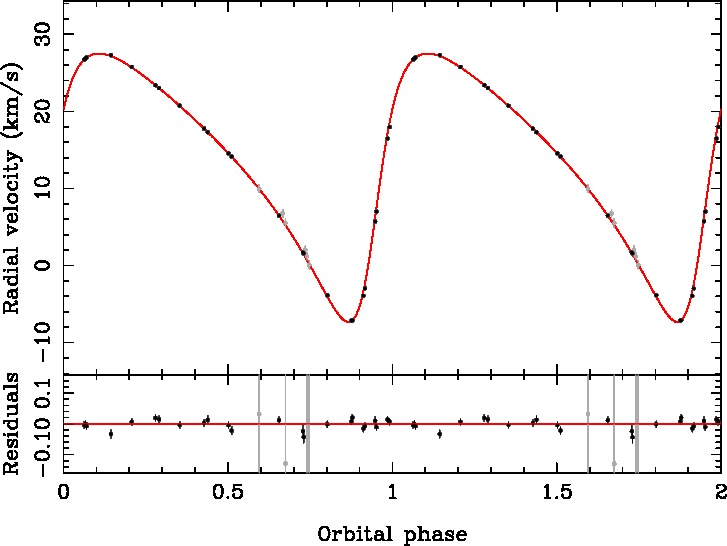}
 \caption{Phase-folded radial velocity plot for the main-sequence star in
   TYC\,7218-934-1. The lower panel shows the residuals to the best fit. Data
   from the Du Pont echelle are shown in grey and from FEROS in black.}
 \label{fig:rvplot}
 \end{center}
\end{figure}

Based on the radial velocities we measured the orbital parameters of TYC\,7218-934-1 
using {\sc exofast}
\citep{eastman13}. The best fit orbit is shown in Fig.~\ref{fig:rvplot} and has a period of 13.6 days with an eccentricity of $e=0.46$. The full orbital solution is detailed in Table~\ref{tab:orbit}. Using the binary mass function, the minimum mass of the companion to the G2V star is 0.2M$_\odot$.


\begin{table}
 \centering
  \caption{Orbital parameters for TYC\,7218-934-1. Parameters highlighted with
  a $^*$ were calculated assuming the G star has a mass of
  $1.04\pm0.02$M$_\odot$.}
  \label{tab:orbit}
  \begin{tabular}{@{}lccc@{}}
  \hline
  Parameter                  & Value         & Error    & Unit     \\
  \hline
  Period                     & 13.601486     & 0.000050 & Days     \\
  Eccentricity               & 0.455         & 0.005    & -        \\
  Argument of periastron     & -113.93       & 0.67     & Degrees  \\
  Time of periastron         & 2456895.86756 & 0.00003  & BJD(TDB) \\
  RV semi-amplitude          & 17.397        & 0.037    & \kms     \\
  Systemic velocity          & 13.385        & 0.010    & \kms     \\
  Semi-major axis$^*$        & 0.113         & 0.001    & au       \\
  Min. companion mass$^*$ & 0.200         & 0.005    & M$_\odot$ \\
  \hline
  \end{tabular}
\end{table}

The eccentric orbit of a binary star that contains a G-type
star and a WD with an orbital period of just 13.6\,days is puzzling.
Such close WD binaries are supposed to form through common envelope evolution,  
in which drag forces within the envelope should quickly circularise 
the orbit. As this quick circularization is a rock-solid prediction of common envelope theories, 
the close eccentric binary in \tycc must have either formed due to the action of 
another mechanism or the WD is not responsible for the velocity variations measured for  
the G2-type secondary star. 

As a non negligible fraction ($\simeq40\%$) of close main sequence binary stars with orbital periods around 13 days are often members of hierarchical triple systems (\citealt{tokovinin2006}, \citealt{tokovinin2014b}), such a configuration could represent a reasonable explanation for the eccentricity observed in \tyc.
This hypothesis, however, leaves two quite distinct possibilities for the
evolutionary history and current configuration of \tyc. 
First, it could indeed be a close WD+AFGK star with an
unseen low-mass companion, where the close inner binary configuration was achieved through the Kozai-Lidov (KL) effect 
generated or enhanced by the mass loss when the progenitor of the WD evolved off the main-sequence  \citep{Shappee&Thompson},
or it might be a close main sequence star binary consisting of the G2V
star and an unseen low-mass companion where the WD is the distant
third object and not part of the inner binary star. In the latter case, the
close inner main sequence binary star might have been affected by 
triple-system dynamics.

\section{The triple nature of \tycc confirmed with SPHERE}

To test the idea of \tycc being a triple system, we observed it with the high contrast imager SPHERE \citep{Beuzit_2019} in the IRDIFS mode. Acquisition of direct imaging was made with IRDIS \citep{2008SPIE.7014E..3LD} in the dual band imaging mode \citep{2010lyot.confE..48V} using the H2-H3 bands with simultaneous spectro-imaging using IFS \citep{2008SPIE.7014E..3EC} in the Y and J spectral bands. The aim of these observations was to confirm the triple nature of the system and to characterise the third object, i.e. to answer the question of whether it is a low mass main sequence star or the WD. 

\subsection{Data reduction}

The SPHERE data were reduced exploiting the data reduction and handling (DRH - \citealt{2008ASPC..394..581P}) pipeline.
The IRDIS data were first preprocessed (sky background subtraction,
flat-fielding, bad-pixels correction). The frames were recentered using the
initial star center exposure with the four satellite spots. The IFS
preprocessing consists of background subtraction and flat-field calibration.
Then, each frame was calibrated with the integral field unit (IFU) flat.  For
wavelength calibration, the IFU was illuminated with four 
monochromatic lasers of known
wavelength. The result of this data preprocessing was a data cube of 39 monochromatic frames. 

After pre-processing, IRDIS and IFS data were reduced with {\it ad hoc}
IDL routines presented in \cite{2014A&A...572A..85Z,2016A&A...587A..57Z} to perform the Angular Differential Imaging
\citep[ADI,][]{2006ApJ...641..556M}. The two IRDIS filters were processed
with the KLIP \citep{2012ApJ...755L..28S} method, as described in detail in
\citet{2016A&A...587A..57Z}. The IFS data were processed using two independent
codes that imply KLIP and a custom principal component analysis (PCA) code, as
presented in  \citet{2016A&A...587A..57Z} and \citet{2015A&A...576A.121M},
respectively.  

For the extraction of the spectrum, we used ADI separately for each individual
IFS channel, since the object is very bright. 
We applied the fake negative planets technique 
\citep[see, e.g.][]{2010Sci...329...57L,2011A&A...528L..15B}, which
is very effective in taking into account the self-subtraction of the flux.  

\subsection{A distant WD companion to a close main sequence binary \label{sec:spherewd}}

\begin{figure}
\begin{center}
 \includegraphics[width=\columnwidth]{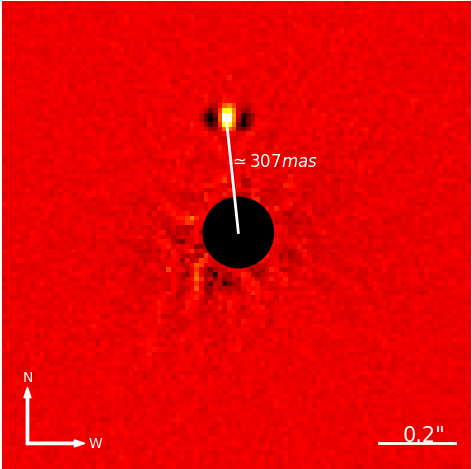}
 \includegraphics[width=\columnwidth]{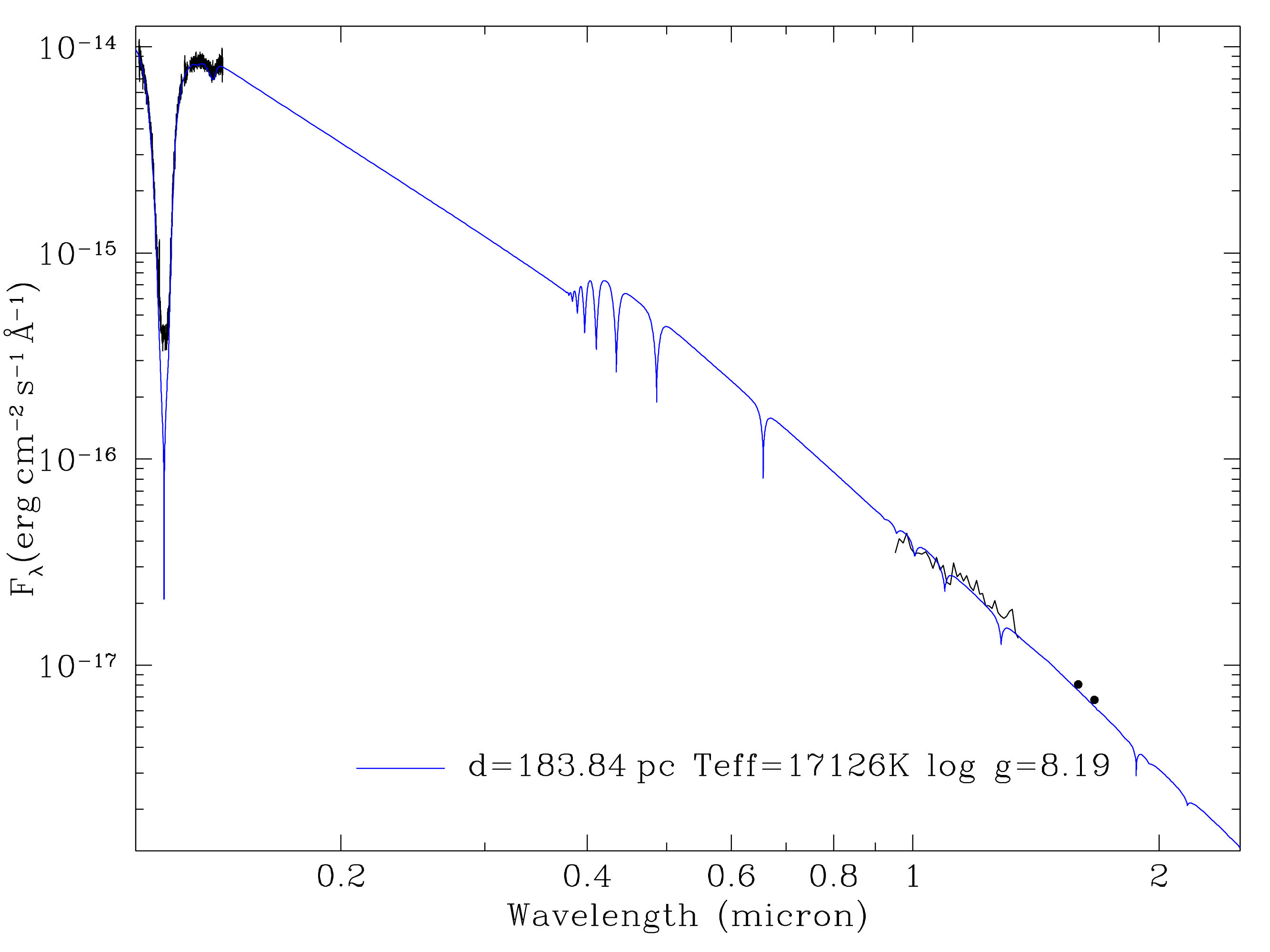}
 \caption{\textbf{Upper panel:} IRDIS image of the third object in the H2 band. The black circle denotes the position of the ALC\_YJH\_S coronograph. The projected separation of the WD relative to the central binary is 307 \textit{mas}.
 \textbf{Bottom panel:} Spectral fit for the HST (black data at $\approx1100$\,\AA ) and SPHERE data (black line at $\approx10000$\,\AA  ) of \tyc. The blue line is the best fit for both HST and SPHERE data at 184\,pc, the distance estimated by the GAIA DR2. The two black points, from left to right, are the H2 and H3 fluxes of the WD respectively, with magnitude differences $\Delta$H2=8.22 $\pm$ 0.03 and $\Delta$H3=8.30 $\pm$ 0.03 with respect to the central binary.}
 \label{fig:sphplot}
 \end{center}
\end{figure}

The SPHERE IFS spectrum (which covers a wavelength range of $9000-10\,400$\,\AA) and the H2 (15930\,\AA) and H3 (16670\,\AA) fluxes are shown in Fig. \ref{fig:sphplot}, along with the COS spectrum of the system, and the best-fit WD model to the COS data (Sect.\,\ref{sec:coswd}). It is evident that the SPHERE observations are entirely consistent with the wide companion in \tycc being the WD, located at a angular distance of 307 \textit{mas}, rather than a low-mass K/M-type star. For completeness, we fitted the COS and SPHERE data together following the same procedure as in Sect.\,\ref{sec:coswd}, and find $\Teff= 17060\pm150$\,K and $\log g = 8.17\pm0.02$, i.e. in agreement with the fit to the far-ultraviolet data alone.

To further support our interpretation of the SPHERE detection, 
we calculate the probability P($\Theta,m$) of chance alignments with background sources within an angular distance $\Theta$. Following \citet{2000AJ....120..950B} we estimate this probability as: 
 
\begin{equation}
  P(\Theta,m_\mathrm{lim})= 1-e^{-\pi \Theta \rho(m_\mathrm{lim})}, 
\end{equation}  

where $\rho(m)$ is the cumulative surface density of background sources down to a limiting magnitude m$_\mathrm{lim}$. In order to calculate $\rho(m_\mathrm{lim})$, we use the Besan\c{c}on galaxy model\footnote{https://model.obs-besancon.fr/modele\_home.php} (\citealt{refId0}) to generate a synthetic \textit{JHK}$_\mathrm{s}$ photometric catalogue of point sources within 0.5 square degrees, centred in the coordinates of \tyc.
The magnitude of the companion is m$_\mathrm{lim}\simeq 18.4$ in the H band, which results in a very low probability  
of $P(\Theta,m_\mathrm{lim})=10^{-4}$ for a background source to be located within $\Theta=307$ \textit{mas} of \tyc. 
We hence conclude that the SPHERE data unambiguously demonstrate that the wide component in the stellar triple \tycc is a $\simeq0.7\,\Msun$ WD.

This implies that the eccentric orbit of the G-type star most likely corresponds to an inner main sequence binary with an unseen companion with a minimum stellar mass of 0.2\Msun (as determined previously from the orbital solution, see Sect. \ref{radial curve}). Assuming that the unseen companion is a main sequence star, we constrain its maximum mass by testing up to which stellar mass we obtain agreement with the archival photometry of \tycc in the G,V,H,J and K$_s$ bands. We used 
VOSA \citep{2008A&A...492..277B} with the Koester (\citealt{koester10-1}) and BT-settl spectra libraries for the WD and the main-sequence stars in the inner binary, respectively. For the unseen companion spectrum we used templates with $\log g=5.0$, [Fe/H]=$0.0$ and effective temperature spanning $\Teff=2500 - 5800$ K with a step size of $\Delta\Teff=500$ K. In total agreement with our interpretation of \tyc, the infrared photometry of the system fits better by including an additional component from the companion that is not seen in the optical spectra. We find acceptable fits (p-value<0.05) for stellar temperatures from 3000 to 5000\,K corresponding to stellar masses of 0.31-0.80\Msun according to the Torres relation (\citealt{torresetal10-1}). The best fit model is obtained assuming a K-type star with $\Teff=4000$ K and mass of 0.55\,M$_\odot$.

\section{The potential evolutionary history of the triple system}

We confirmed the triple nature of \tyc, identifying the WD 
as the third object at a projected separation of $\approx 56$\,au 
(assuming a distance to the system of 184 pc) from the inner binary. 
This inner binary most likely 
consists of the G2V star and an (in the optical) unseen K-M dwarf main sequence companion. 

In order to investigate the possible evolutionary history of this triple system,
we estimated whether it is likely that the observed inner eccentric orbit in 
\tycc was caused by Kozai-Lidov mechanisms (KLMs), as close binaries with 
short orbital periods likely have tertiary companions \citep{tokovinin2006}. 
To evaluate the current presence of eccentricity oscillations in the system due 
to the standard and eccentric KLMs \citep{Lithwick&naoz} 
we compare their time-scales with the time-scale of general relativity precession, which can detune the eccentricity oscillations produced by the KLMs. According 
to \citet{li2015} the time-scales of the standard KLM, eccentric KLM and general relativity precession are 

\begin{equation}
t_\mathrm{SKLM}=\frac{2 \pi a_\mathrm{out}^3 (1-e_\mathrm{out}^2)^{3/2} \sqrt{(M_1+M_2)(1-e_\mathrm{in}^2)} }{G^{1/2}a_\mathrm{in}^{3/2}M_3},
\label{eq:1}
\end{equation}

\begin{equation}
t_{\mathrm{EKLM}}=\frac{t_\mathrm{SKM}}{\epsilon_\mathrm{oct}}
\label{eq:2}
\end{equation}    

\noindent
and
\begin{equation}
t_\mathrm{GRP}=2\pi \frac{a_\mathrm{in}^{5/2}c^{2}(1-e_{1}^{2})}{3G^{3/2}(M_1 +M_2)^{3/2}}
\label{eq:3}
\end{equation}
%
respectively, where $a$ is the semi-major axis, $e$ the eccentricity, $M_1$ and $M_2$ 
the masses of the stars in the inner binary and $M_3$ the mass of the third star, while
the subscripts $in$ and $out$ refer to the inner and outer orbit. The term
\begin{equation}
\epsilon_{\mathrm{oct}}=\bigg|\frac{M_1-M_2}{M_1+M_2}\bigg| \left( \frac{a_\mathrm{in}}{a_\mathrm{out}}\right) \frac{e_\mathrm{out}}{1-e_\mathrm{out}^2}
\label{eq:4}
\end{equation}
%
measures the impact of the eccentric KLM on the orbital evolution
of the system relative to the standard KLM. 
For $\epsilon_\mathrm{oct}\gtrsim 0.01$ the eccentricity oscillations due to the eccentric KLM become 
important \citep{Lithwick&naoz}. Thus, if $t_\mathrm{SKM}>t_\mathrm{GRP}$ or $t_\mathrm{EKM}>t_\mathrm{GRP}$ Kozai-Lidov oscillations 
are suppressed by general relativity precession. 
To evaluate numerically Eqs. \ref{eq:1}-\ref{eq:4} we assumed the projected separation we derived from the SPHERE observations to be the real separation of the third object at apastron, which represents a lower limit of the true value of $a_{out}$. As the time-scale of Kozai-Lidov oscillations decreases with the separation of the third star, this implies that, if in this assumed configuration KLMs are not important, they are clearly not important in \tycc right
now. As the eccentricity of the orbit of the third object is unknown, we leave it as a free parameter.

\begin{table}
 \centering
  \caption{Current parameters for TYC\,7218-934-1. The outer orbit is only constrained by the projected separation of the tertiary detected with SPHERE. For the mass of the companion to the G star in the inner binary we can only derive a lower limit. }
  \label{tab:cpars}
  \begin{tabular}{@{}lccc@{}}
  \hline
  Parameter                  & Value          & Unit      \\
  \hline
  G star mass                & 1.04$\pm$0.02           & M$_\odot$ \\
  Companion mass             & $0.2\lappr$M$\lappr0.8$          & M$_\odot$ \\
  WD mass                    & $0.73\pm0.03$           & M$_\odot$ \\ 
  Inner Period               & 13.601486$\pm0.00005$     & d      \\
  Outer projected separation       & 56             & au      \\
  Outer period               & $\gappr10^{4.8}$     & d	  \\
  Eccentricity               & 0.455$\pm0.005$          & -         \\
  
  \hline
  \end{tabular}
\end{table}

\begin{figure}
    \centering
    \subfloat{
     \includegraphics[width=\columnwidth]{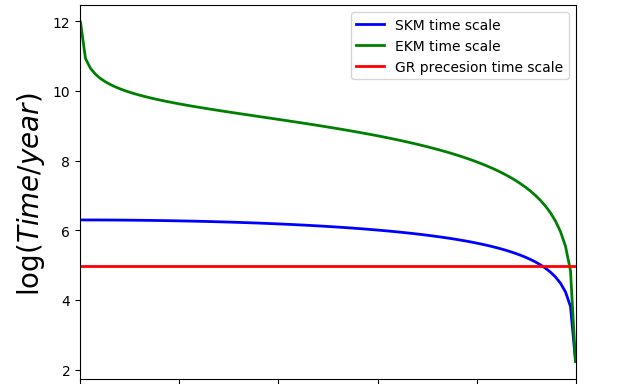}
    }
    \vspace{-3 mm}
    \subfloat{
      \includegraphics[width=\columnwidth]{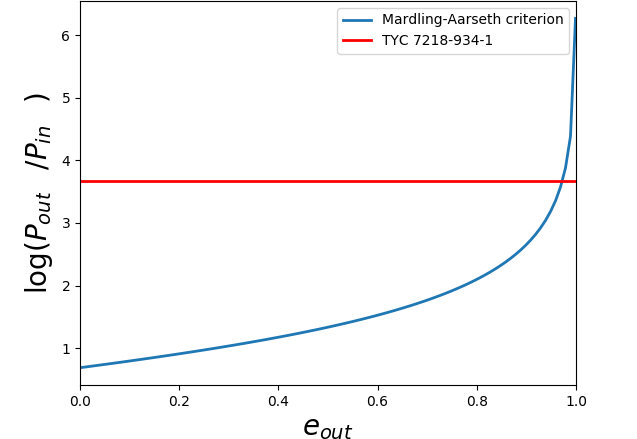}
    }
    \caption{\textbf{Upper panel:} Comparison between the Kozai-Lidov mechanisms time-scales (blue and green lines) and the general relativity precession time-scale (red line) for the \tycc configuration given in table \ref{tab:cpars}, leaving the outer eccentricity $e_\mathrm{out}$ as a free parameter. Only for values of $e_\mathrm{out}$ greater than $\approx0.93$ and $\approx0.98$ the standard and eccentric Kozai-Lidov oscillations are present. \textbf{Bottom panel:} Using the stability criterion derived by \citet{mardling} for the maximum ratio between inner and outer period (blue line),
    the highest eccentricity allowed for \tyc, given its current minimum period ratio (red line), is $\approx 0.97$.}
    \label{fig:my_label}
\end{figure}

Using the system parameters summarized in Table \ref{tab:cpars}, 
we find that KLMs could only be important for very high eccentricities of the 
outer orbit, exceeding $\approx 0.93$ (see upper panel of Fig. \ref{fig:my_label}). However, such high eccentricities
would bring the system very close to being unstable despite the rather large difference 
in separation between the two orbits, as shown in the bottom panel of Fig. \ref{fig:my_label}. 
We therefore conclude that it is very unlikely that the system is currently affected by KLMs. However, the fact that KLMs are currently not important does not imply that the evolution of the system was always unaffected by this mechanisms. 
If KLMs plus tidal forces affected the inner orbit, it could have been much larger in the past, i.e. the current period (13.6\,days) represents only a lower limit on the initial inner orbital period. 
Reconstructing the exact history of \tycc is of course impossible, as the outer orbit is poorly constrained. However, the parameters we derived for \tycc fit well into 
a scenario for triple evolution recently suggested by \citet{moe&kratter}. 


These authors propose that two different regimes exist for tidal effects in binary stars. 
For large eccentricities, strong dynamical tides dominate while for smaller eccentricities only weak tidal forces are expected.  
The eccentricity at which the transition between these two regimes occurs,  
the so-called transition eccentricity $e_{\mathrm{trans}}$, 
has been considered as a free parameter by \citet{moe&kratter}. 
However, they showed that setting $e_{\mathrm{trans}}=0.5$ would explain the observed population of main-sequence triple systems with inner periods of less than 6 days. 
Interestingly, in the context of \tyc, according to this scenario, inner binaries should concentrate close to $e_{\mathrm{trans}}$ when the orbital period is above the circularization period which is P$_{\mathrm{circ}}=11$ days for solar-type MS binaries \citep{Raghavan_2010}. 
Therefore, with its inner eccentricity of 0.46 and its inner period of 13.6 days, 
\tycc agrees very well with the scenario outlined by \citet{moe&kratter}. 

Thus, a plausible scenario for the evolution of \tycc can be described as follows: 
the initial configuration consisted of an inner binary with $P_{\mathrm{in},0} > 100\,\mathrm{d}$ days affected by KL oscillations. The large eccentricities generated led to strong dynamical tides decreasing the orbital period and eccentricity of the inner binary. When the eccentricity became smaller than $e_{\mathrm{trans}}=0.5$, 
the inner eccentricity began to only very slowly decrease 
since weak-friction equilibrium-tide evolution is a slow process.
When the eccentricity was close to its current value (0.455),
the tertiary evolved into the WD we observe today. 
The corresponding mass loss increased the outer period and turned off KLMs. In the future, the inner binary will continue to slowly circularise.

\section{How many triples do we expect to hide among the WD/AFGK targets?}

As \tycc has been identified in our survey of post common envelope binaries (PCEBs) consisting of WD+AFGK binaries, an urgent question is of course, how many more triples with a K-M dwarf and a WD, being each one either the tertiary component or the AFGK companion, might contaminate our sample. To roughly evaluate the possible contamination due to triple systems containing a WD and a K-M dwarf, we used the algorithm presented by \citealt{tokovinin2014b} and generated a population of binary and triple systems composed of zero-age main sequence stars. As the algorithm only returns orbital periods and mass ratios, parameters like the eccentricities and inclination between the inner and outer orbits are assumed to follow the distributions used in \citealt{fabrycky_and_tremaine_2007}.
Once the initial population is generated, we assign a born-time to each system from a uniform distribution between 0 and 10 Gyr, and select only those where one star (the most massive one of the system) had enough time to evolve into a WD. 
We then use the \textit{Binary Stellar Evolution} (BSE) code \citep{hurley2002} to evolve those systems that enter a common envelope phase when the more massive star of the binary (or the inner binary of a triple) evolves off the main sequence. 
For triples in which the most massive star evolves into a WD without interacting with the other stars in the system, we simply use the the adiabatic mass loss model to determine the effect on the orbital separations.

Based on this generated population of PCEBs and triples with configurations AFGK/WD+KM (the WD as part of the inner binary) and AFGK/KM+WD (the WD being the tertiary), we 
derive the upper and lower limit of the fraction of the triple system contamination. 
First, we obtain a lower limit on the fraction of triples in our sample by comparing the pure number of PCEBs (with AFGK secondary 
stars) with the number of triple systems with a WD tertiary (AFGK/KM+WD configuration) and with the inner binary having an orbital period shorter than 100 days, ignoring any possible triple dynamics.  
We find that PCEBs are roughly a hundred times more frequent than triples with white dwarfs tertiares. This value could be slightly higher ($\sim2\%$) as the simulator of \citealt{tokovinin2014b} does not fully reproduce the fraction of inner sub-systems 
with periods less than 100 days. 
In order to complement this lower limit with an upper limit, 
we derived the number of systems with an initial inner period longer than 100 days that are affected by  
KLMs either before or after the evolution of the most massive star into a WD. We find that the fraction of contaminants due to triple systems with configurations AFGK/WD+KM and AFGK/KM+WD increases to 15 per cent. We thus conclude that the contamination by triple systems of our sample is most likely in the range of 1-15 per cent.


To estimate this upper and lower limit 
we used the algorithm of \citealt{tokovinin2014b} and the distributions in \citealt{fabrycky_and_tremaine_2007}, which 
have been derived from observations of field stars with masses in the range $0.8\lesssim$M$\lesssim 1.3$ M$_{\odot}$.  
This represents a reasonable approach although it contains rather rough approximations. 
First, we used these distributions to create a population of
zero-age main sequence binary and triple systems while they have been 
derived from observations of field stars. 
Triple stars in the field, however, may already be
affected by evolutionary effects. These evolutionary effects might have led to a reduction of the inner period due to tidal effects which somewhat softens our lower limit. 
Second, we used the distributions for stellar components beyond the mass range of the observed sample. This approach should be correct in terms of the mass ratio and period distributions as those of intermediate and high mass stars do not differ significantly from those used for sun-like stars \citep{Duchene_Kraus2013}. 
The multiplicity fractions, however, may significantly increase as a function of primary star mass. For example, nearly all B type stars seem to be members of multiple systems, with a triple fraction of $\sim 50\%$ \citep{toonen2016,moe&kratter}. 
These larger multiplicity fractions for massive progenitor systems might somewhat soften our upper limit. 

\section{Summary and Conclusions}
We report the discovery of the first hierarchical triple system in the \textit{White Dwarf Binary Pathways Survey}. TYC 7218-934-1, initially identified as a G-type star orbited by a WD with a period of $P_\mathrm{in}=$13.6 days, showed an eccentric orbit, being inconsistent with the circularization process of binaries that passed through the common envelope phase. 
Based on the hypothesis that the object is therefore rather a triple system than a post common envelope binary 
we observed \tycc with the VLT/SPHERE in the IRDIFS mode to look for the third star. 
The performed observations confirmed the presence of a third object at a projected separation of $\approx$56 AU from the inner binary, and the IFS spectrum in the YJ-band confirmed that this tertiary corresponds to the WD. 
We then performed binary and triple population synthesis to 
estimate the fraction of triple systems such as \tyc that contaminate our sample 
of post common envelope WD+AFGK binaries. 
We found that $\approx$1-15 per cent of the observed close WD+AFGK binaries could be triple systems with a WD tertiary. This fraction of contaminating triple systems is clearly acceptable and therefore the project remains a 
promising pathway towards progressing with the SN\,Ia progenitor puzzle.

\section*{Acknowledgements}
MRS acknowledges financial support from FONDECYT (grant number 1181404).
SGP acknowledges the support of the STFC Ernest Rutherford Fellowship.
AZ acknowledges support from the CONICYT/PAI (grant PAI77170087).
BTG was supported by the UK STFC grant ST/P000495.
RB acknowledges support from FONDECYT Post-doctoral Fellowship Project 3180246, and from the Millennium. Institute of Astrophysics (MAS).
CC acknowledges support from DGI-UNAB project DI-11-19/R.
MSH thanks Doctorado Nacional CONICYT 2017 folio 21170070.
CT acknowledges financial support from Fondecyt No. 1170566.
MZ acknowledges support from CONICYT/PAI (grant 79170121) and CONICYT/FONDECYT (grant 11170559). We also thank 
the referee Steve B. Howell for helpful suggestions 
that improved the presentation of this paper.  
%



\bibliographystyle{mn2e}
\bibliography{aabibtyc} 



\bsp	
\label{lastpage}
\end{document}